\documentclass[conference]{IEEEtran}
\usepackage{cite}

\ifCLASSINFOpdf
  \usepackage[pdftex]{graphicx}
  \graphicspath{{img/}}
  \DeclareGraphicsExtensions{.pdf,.jpeg,.png}
\else
  \usepackage[dvips]{graphicx}
  \graphicspath{{img/}}
  \DeclareGraphicsExtensions{.eps}
\fi
\usepackage[cmex10]{amsmath}
\usepackage{algorithmic}

\usepackage{array}

\usepackage{mdwmath}
\usepackage{mdwtab}
\usepackage[caption=false,font=footnotesize]{subfig}
\usepackage{fixltx2e}
\usepackage{url}

\usepackage{fancyhdr}
\begin{document}
\title{Enhancing Fault Tolerance Capabilities\\in Quorum-based Cycle Routing \\ \footnotesize The final publication is available at IEEE via \url{http://dx.doi.org/10.1109/RNDM.2015.7324305}}
\author{\IEEEauthorblockN{Cory J. Kleinheksel, Member, IEEE and Arun K. Somani, Fellow, IEEE}
	\IEEEauthorblockA{Electrical and Computer Engineering, Iowa State University, Ames, Iowa 50011\\
		Telephone: (515) 294--0442, Fax: (515) 294--9273, Email: \{cklein, arun\}@iastate.edu}}
\maketitle

\begin{abstract}
	In this paper we propose a generalized $R$ redundancy cycle technique that provides optical networks almost fault-tolerant communications. More importantly, when applied using only single cycles rather than the standard paired cycles, the generalized $R$ redundancy technique is shown to almost halve the necessary light-trail resources in the network while maintaining the fault-tolerance and dependability expected from cycle-based routing.
	
	For efficiency and distributed control, it is common in distributed systems and algorithms to group nodes into intersecting sets referred to as quorum sets.  Optimal communication quorum sets forming optical cycles based on light-trails have been shown to flexibly and efficiently route both point-to-point and multipoint-to-multipoint traffic requests. Commonly cycle routing techniques will use pairs of cycles to achieve both routing and fault-tolerance, which uses substantial resources and creates the potential for underutilization.  Instead, we intentionally utilize $R$ redundancy within the quorum cycles for fault-tolerance such that every point-to-point communication pairs occur in at least $R$ cycles.  The result is a set of $R=3$ redundant cycles with 93.23 - 99.34\% fault coverage even with two simultaneous faults all while using 38.85 - 42.39\% fewer resources.
\end{abstract}

\section{Introduction}
We developed a novel method to deliver the almost fault-tolerant capabilities of cycles in an optical network and the potential for significantly reducing the resource utilization when compared to the state-of-art techniques. Cycle-based routing can satisfy both dynamic point-to-point and multi-point optical communications. Cycles are created using quorums of nodes. Within a cycle, multicasts to all nodes in that cycle is possible. The quorum intersection property and the use of cyclic quorums sets provide all of the unicast capabilities.  Exploiting the same properties we can achieve efficient broadcasts with $O(\sqrt{N})$ multicasts.

Optical networks are depended upon for high speed communications in distributed algorithms, as much as they are needed for the arbitrary point-to-point communications.  Failures within a network are to be expected and can happen as much as every couple days.  Protecting against these optical circuit faults is critical and there are many different approaches depending on the network needs and individual circumstances.

For efficiency and distributed control, it is common in distributed systems and algorithms to group nodes into intersecting sets referred to as quorum sets.  Quorums sets for cycle-based routing to efficiently support arbitrary point-to-point and multi-point optical communication were first proposed in \cite{dlastine2014quorum} with fault-tolerance analyzed in \cite{ckleinheksel2015quorum}.  In this paper we apply the established quorum set theory and generalize additional requirements proposed in \cite{ckleinheksel2015redundant} to form suitable $R$ redundant quorums for our optical network routing.

The rest of the paper is organized as follows.  Sections \ref{sec:NetworkModel}, \ref{sec:light_trailDefinition}, and \ref{sec:cycleRouting} establish the network model, node communication, and path routing / fault-tolerance.  In Section \ref{sec:Quorums}, we discuss our application of the distributed efficiency of the quorum sets to routing optical cycles.  We also define and generalize $R$ redundant quorums sets for this application.  Section \ref{sec:pairedCycleAnalysis} analyzes the performance and fault-tolerance of our redundant quorums set cycle routing techniques for the state of art paired cycle routing solutions.  Lastly, Section \ref{sec:singleCycleAnalysis} analyzes our proposed single cycle routing solutions and compares them to the more resource intensive paired cycle routing solution.

\section{Network Model} 
\label{sec:NetworkModel}
No two fiber-optic networks are the same.  Some stretch hundreds of kilometers, while other networks are contained within buildings or rooms.  Regardless of the physical environment, these optical circuits are depended upon for high speed communications.  Thus it is important to extract the network's critical components that affect its ability to deliver reliable, arbitrary point-to-point and multi-point communications.

These fiber-optic networks consist of several transmitters and receivers interconnected by fiber-optic cables.  As you might expect, transmitters and receivers are typically found together and can generically be called an optical node.  The cables form the links (i.e. edges) between those nodes, which leads to a convenient model of a network in terms of a graph $G = (V,E)$.  $V$ is the set of nodes in the network and $E$ is the set of edges. 

Edge $(a_i,a_j)$ is a fiber-optic link connecting nodes $a_i$ and $a_j$ in the network, where $a_i,a_j \in V$ and $(a_i,a_j) \in E$.  It is a general assumption that the same set of optical wavelengths are available on all edges in $E$.  The number of wavelengths available per optical fiber is dependent on the fiber-optic cables and the transmitter/receiver pairs.

\section{Light-trails} 
\label{sec:light_trailDefinition}
Lightpaths were a critical building block in the first optical communications, but required significant traffic engineering and were not able to support multicast traffic.  Light-trails were proposed in \cite{chlamtac2003light} as a solution to these challenges and could be built using commercial off-the-shelf technology.  

Light-trails enable fast, dynamic creation of an unidirectional optical communication channel.  This communication channel allows for channel receive and transmit access to all connected nodes, making them more suitable for IP centric traffic \cite{fang2004optimal}.  Point-to-point communications from an upstream node to a downstream node can be scheduled on the shared light-trail.  Similarly, an upstream node can multicast to any number of downstream nodes.  

\begin{figure}[t]
	\centering
	\includegraphics[width=2.5in]{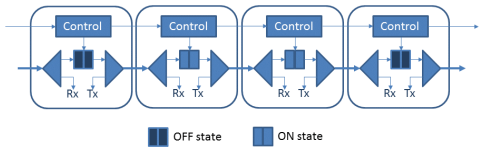}
	\caption{Four nodes in a light-trail architecture.}
	\label{fig:lighttrail_arch}
\end{figure}

An example four-node light-trail can be seen in Fig. \ref{fig:lighttrail_arch}.  Optical shutters isolate an optical signal to a specific light-trail and allow for wavelength reuse within the network. Start and end nodes have their optical shutters in the \textit{off} state, while intermediate nodes have their optical shutters in the \textit{on} state.  The communication is all optical from the start node to the end node.  Being all optical avoids any energy inefficiencies and time delays associated with unnecessary Optical-to-Electrical-to-Optical (O/E/O) conversions at intermediate hops.

Nodes can receive from the incoming signal while the signal is simultaneously continuing to downstream nodes, sometimes referred to as a drop and continue function.  Early technology supported only a few wavelengths, however the latest devices may support over 100 channels, hence allowing multiple light-trails to share the same edge in the network for a combined over 1-Terabits/s.  

A scheduling protocol is in place to avoid collisions within a light-trail and controls when nodes are able to transmit to downstream nodes.  The scheduling is generally assumed to occur over a control channel, which may or may not be separate from the shared optical fiber that is being used for the light-trail.

\section{Light-Trail, Cycle Routing, and Fault-Tolerance}
\label{sec:cycleRouting}
Point-to-point and multi-point traffic requests have a set of nodes $C = \{a_i,...,a_j\}$ that wish to communicate.  Failures within an optical network are to be expected, hence requests need to be protected against network faults.  Establishing a primary and backup multicast path from every node to every other node in $C$ can be a waste of resources though.  

The generalization of p-cycle protection to allow for path and link protection was proposed by \cite{grover2003extending}.  P-cycle protection of unicast and multicast traffic networks requires preconfiguration and the offline nature allows for the efficient cycles to be selected \cite{zhang2008performance,zhang2008optimizations}.  
The use of path-pair protection, linked-based shared protection, spanning paths, and p-cycles to protect multicast sessions have all been proposed for WDM networks as well \cite{singhal2003provisioning,qing2005protecting,luo2006protecting,zhang2007applying,feng2008intelligent}.

The Optimized Collapsed Rings (OCR) single link protection heuristic was developed to address the heterogeneous, part multicast / part unicast, nature of WDM traffic\cite{khalil2005pre}.  The Multi-point Cycle Routing Algorithm (MCRA) uses bidirectional cycles for fault-tolerance and is capable of supporting SONET rings and p-cycles\cite{dlastine2012fault}.  Although finding the smallest cycle supporting the multi-point communication is NP-Complete, the authors were able to show that their heuristic performed within 1.2 times of the optimal cycle size.  ECBRA is a significant improvement of MCRA and outperforms the OCR heuristic \cite{dlastine2011ECBRA}.   

The ECBRA heuristic balances optimality and speed, taking $O(\left|E\right|\left|C\right|^3)$ steps to find a close to optimal cycle.  First, a modified breadth first search is performed on each node in set $C$ of required communication nodes.  The goal is to find a shortest path in $G$ that also has the best ratio of nodes from set $C$ vs. total nodes on the path.  To complete cycle using the chosen shortest path, a path from the sink node returning to the source node must be found.  No links may be used twice.  If all nodes in $C$ are in the cycle, then the cycle search is complete.  If needed, the final step iteratively removes edges from the cycle and inserts paths through missing nodes in $C$.  Because insertion of the node can be cheaper by removing certain links from the cycle rather than others, the optimal edge removal from the cycle and path replacement is computed for each missing node insertion.

\begin{figure}[t]
	\centering
	\includegraphics[width=2.5in]{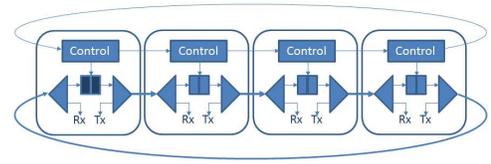}
	\caption{Cycle formed using the light-trail architecture.}
	\label{fig:lighttrail_cycle}
\end{figure} 

In this work, we utilize the light-trail architecture in the form of a cycle.  Figure \ref{fig:lighttrail_cycle} is simply a light-trail where the start and end node is the same node, referred to as the \textit{hub node} of the cycle.  The hub node has its optical shutters in the \textit{off} state, while intermediate nodes have their optical shutters in the \textit{on} state.  The resources at each hub node can be utilized to allow all-to-all communication on the cycle using only one light-trail.  Traffic from a node to nodes downstream requires a single transmission.  Traffic from a node to an upstream node must undergo Optical-to-Electrical-to-Optical (O/E/O) conversion at the hub node and be transmitted on the light-trail a second time.

Alternately, the state of art solutions setup two light-trails, one in each direction.  This enables upstream communications without the energy inefficiencies and time delays associated with O/E/O conversions.  


\section{Quorums} 
\label{sec:Quorums}
In distributed communication and algorithms, coordination, mutual exclusion, and consensus implementations have grouped $N$ nodes into small sets called quorums.  This organization of nodes can minimize communications in operations like negotiating access to a global resource. 

A quorums set minimally has the property that all quorums in the set must intersect.  For distributed implementations, it can also be desirable that each node have equal work and equal responsibility \cite{maekawa1985algorithm}.  

Not every grouping of nodes into sets (quorums) will result in having these three properties, nor will the quorum sizes be minimal.  \cite{maekawa1985algorithm} proved the lower bound on the size of quorums set having these three properties.  Cyclic quorums sets with these properties, like other quorums sets in general, are difficult to find and require an exhaustive search \cite{maekawa1985algorithm,luk1997two}.  Quorums sets were shown to efficiently support arbitrary point-to-point and multi-point communication for cycle-based routing in \cite{ckleinheksel2015quorum,dlastine2014quorum}.  Redundant quorum sets significantly reduced the required amount of network resources while still maintaining a similar level of fault-tolerance\cite{ckleinheksel2015redundant}.

\subsection{Defining Quorums Set}
\label{sec:Quorums:sub:Definition}
$A$ is a set of $N = \left|V\right|$ nodes.  A set $S_i$ is a subset of $A$.  When set $Q$ of subsets (Eq. \ref{eq:setQ}) covers all nodes in $A$ (Eq. \ref{eq:quorumUnion}) and all subsets also have non-empty intersections (Eq. \ref{eq:quorumIntersection}), then set $Q$ is called a quorums set.
\begin{align}
A& = \{a_1,...,a_N\}\\
Q& = \{S_1,...,S_N\}\label{eq:setQ}\\
\bigcup _{i=1}^NS_i& = \{a_1,...,a_N\} = A\label{eq:quorumUnion}\\
S_i \cap S_j& \neq \emptyset, \forall i,j \in 1,2,...,N\label{eq:quorumIntersection}
\end{align}

The lower bounds for the maximum individual quorum size (i.e. $\left|S_i\right|$) in a minimum set is $K$, where Equation \ref{eq:quorumSize} holds and $(K-1)$ is a power of a prime, proved through equivalence to finding a finite projective plane\cite{maekawa1985algorithm}. Additionally it is desirable that each quorum $S_i$ in the quorum set be of equal size (Eq. \ref{eq:equalWork}), such that there is equal work and it is desirable that each node be contained in the same number of quorums (Eq. \ref{eq:equalResponsibility}), such that there is equal responsibility.
\begin{align}
N \leq K&(K-1) + 1\label{eq:quorumSize}\\
\left|S_i\right| = K&, \forall i \in 1, 2, ..., N\label{eq:equalWork}\\
a_i \textrm{ is contained in } K& \: S_j\!\,'s, \forall i \in 1, 2, ..., N\label{eq:equalResponsibility}
\end{align}

Cyclic quorums adhere to these properties \cite{luk1997two}.  Cyclic quorums are unique in that once the first quorum (Eq. \ref{eq:firstquorum}) is defined the remaining quorums can be generated via incrementing  (modulus not shown in Eq. \ref{eq:incrementquorum} for conciseness.)  For simplicity assume $a_1 \in S_1$ without loss of generality (any one-to-one re-mapping of entity ids can result in this assumption.)
\begin{align}
S_1& = \{a_1, ..., a_j\}\label{eq:firstquorum}\\
S_i& = \{a_{1+(i-1)}, ..., a_{j+(i-1)}\}\label{eq:incrementquorum}
\end{align}

\subsection{Redundant Cyclic Quorums Sets}
\label{sec:Quorums:sub:Redundant}
In this section we define and generalize $R$ redundant quorums sets.  The quorum-based cycle solution uses quorums to form a set of communication cycles which were shown to support almost fault-tolerance to fiber optic networks\cite{dlastine2014quorum,ckleinheksel2015quorum}.  As defined in Section \ref{sec:Quorums:sub:Definition}, there are $N$ quorums in a set and each quorum has $K$ nodes.  In analysis of networking capabilities we are interested in whether every node can communicate with every other node.  
Equation \ref{eq:numberPairs} considers the number of communication pairs within a quorum, i.e. the pairs made between $K$ nodes communicating with $(K-1)$ other nodes in a single quorum.  
Equation \ref{eq:totalNumberPairs} is the total pairs in the size $N$ quorums set solution.  For convenience we set $M$ to be the total pairs for a given network with $N$ nodes and $K$ optimal quorum size.

\begin{align}
\frac{K(K-1)}{2}&=O(K^2)\label{eq:numberPairs}\\
N\frac{K(K-1)}{2}&=O(NK^2)\label{eq:totalNumberPairs}\\
M&=O(NK^2)
\end{align}

When the quorum size, $K$, is minimal or larger, every pair of nodes $(a_i,a_j)$ occurs together within a quorum in the set at least once.  Optical networking however requires all directional point-to-point pairs to exist, i.e. both pairs $(a_i,a_j)$ and $(a_j,a_i)$.  Previously this had been addressed by pairing each cycle with the same cycle with its direction reversed.  The key change in \cite{ckleinheksel2015redundant} is an added requirement that every pair $(a_i,a_j)$ would occur together within at least two quorums rather than just one, which sought to eliminate the need for paired cycles.  The number of quorums in the solution remained the same $N$, hence to create the additional pairs the quorum size had to be enlarged to $\hat{K}$.  Equation \ref{eq:totalR2NumberPairs} calculates the number of node pairs in quorums of size $\hat{K}$.  Equation \ref{eq:R2twiceR1} is our requirement that the total number of pairs have doubled from the original total pairs, $M$.  Finally, Eq. \ref{eq:R2size} solves for size $\hat{K}$ in relation to optimal $K$.
\begin{align}
N\frac{\hat{K}(\hat{K}-1)}{2}&=O(N\hat{K}^2)\label{eq:totalR2NumberPairs}\\
O(N\hat{K}^2)&=2M\label{eq:R2twiceR1}\\
\hat{K}&\approx\sqrt{2}K\label{eq:R2size}
\end{align}

This result is powerful because with a $\sqrt{2}$ factor increase in $K$ the need for paired cycles is reduced and opens the door for considerable resource savings.  Still there are applications that may benefit from improved fault tolerance, hence we further generalize this approach for a generic desired $R$ redundant factor to offer an opportunity to enhance the fault tolerance of our quorum-based cycle routing solution.  Equation \ref{eq:RtimesR1} balances the enlarged quorum size $\hat{K}$ solution against the known $R$ times the optimal solution.  Equation \ref{eq:Rsize} solves for $\hat{K}$ in terms of known $K$.
\begin{align}
O(N\hat{K}^2)&=RM\label{eq:RtimesR1}\\
\hat{K}&\approx\sqrt{R}K\label{eq:Rsize}
\end{align}

To the best of our knowledge, no efficient algorithm is known to find quorums of minimum size, particularly with the additional requirement that entity pairs appear a minimum $R$ times within the quorums set solution.  \cite{luk1997two} used a brute force search to find optimal cyclic quorums for $N=4\dots111$.  Using our generalized result from Eq. \ref{eq:Rsize}, we too used a brute force search beginning with the smallest possible quorum size for a given number of nodes $N$ and a given desired redundancy factor $R$.

The resulting redundant quorums were utilized in the following sections, as we analyzed and enhanced the efficacy of quorum-based cycle routing in optical networking.

\section{Paired Cycles Network Analysis}
\label{sec:pairedCycleAnalysis}
\begin{figure*}[t]
	\centering
	\subfloat[]{\includegraphics[width=1.75in]{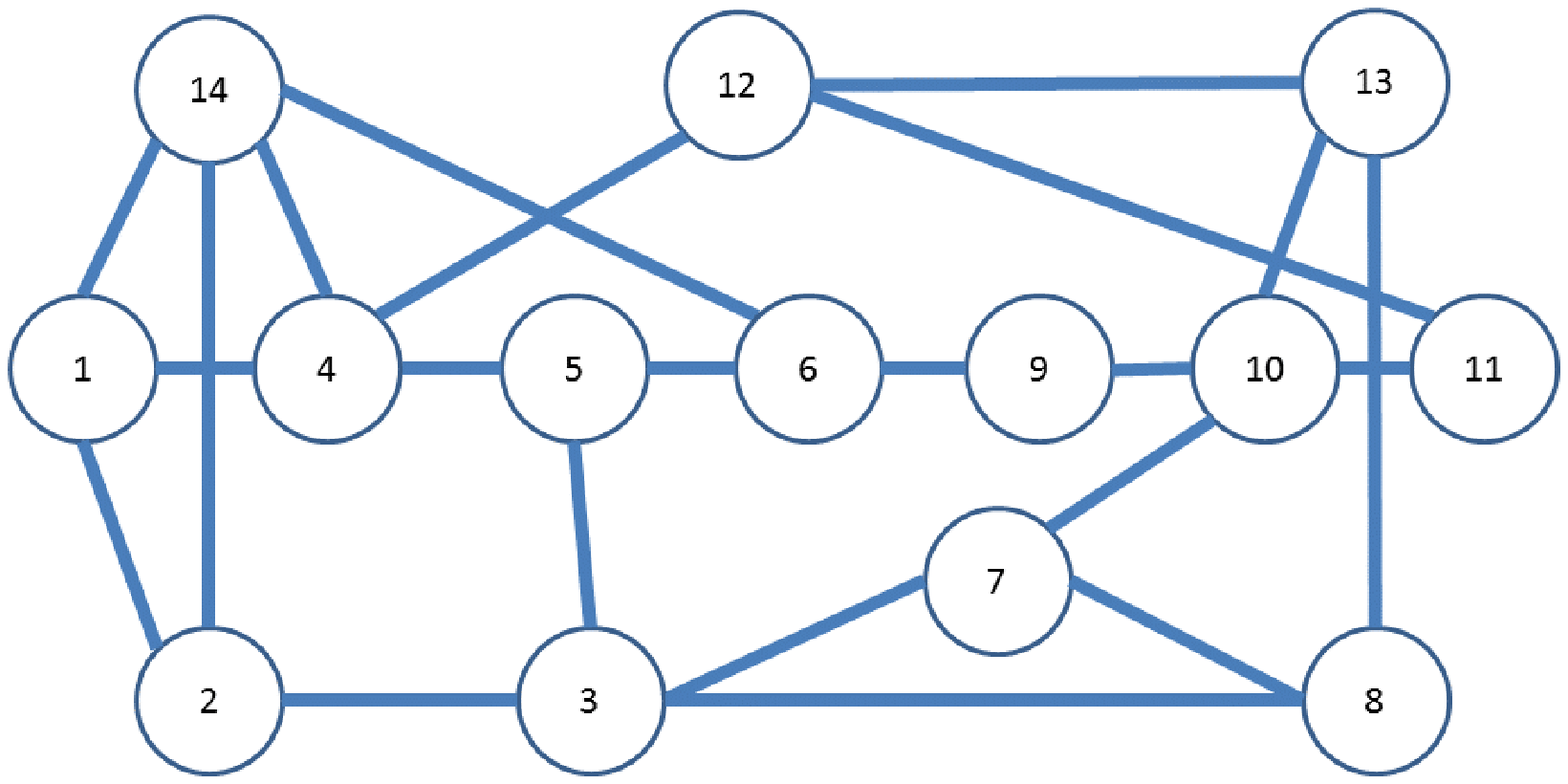}
		\label{fig:Networks:sub:nsfnet}}
	\hfil
	\subfloat[]{\includegraphics[width=1.75in]{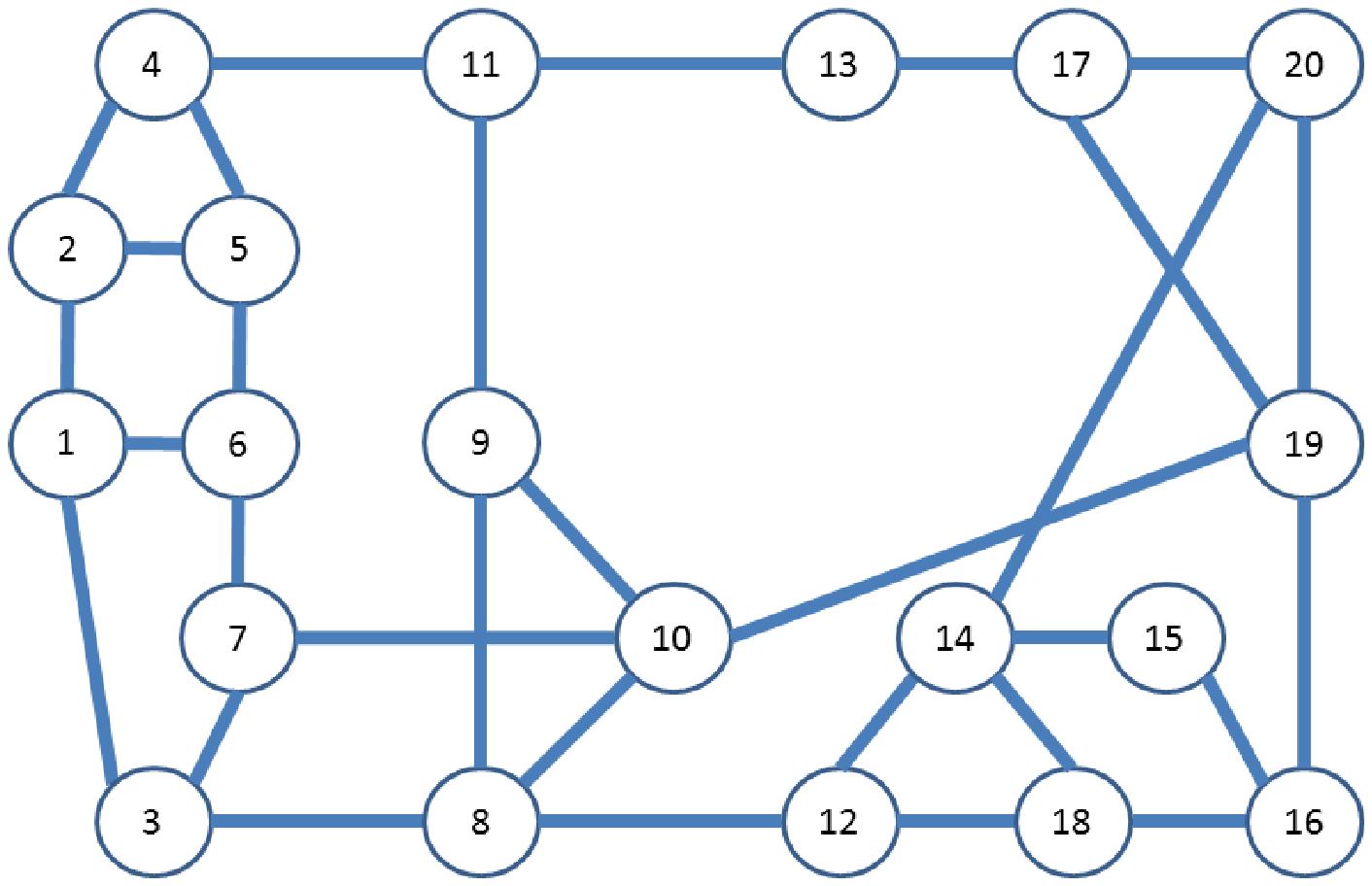}
		\label{fig:Networks:sub:arpanet}}
	\hfil
	\subfloat[]{\includegraphics[width=1.75in]{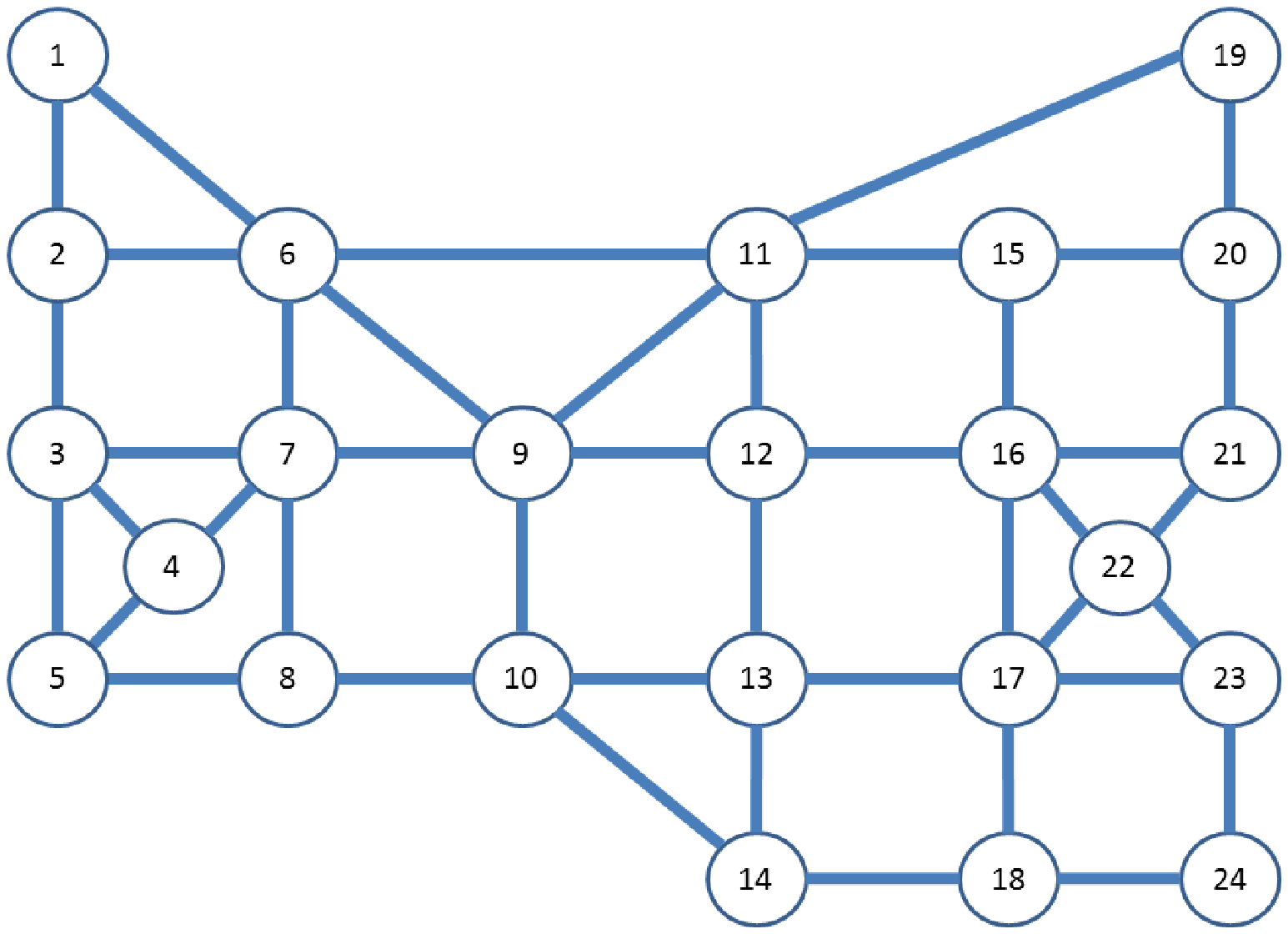}
		\label{fig:Networks:sub:american}}
	\hfil
	\subfloat[]{\includegraphics[width=1.75in]{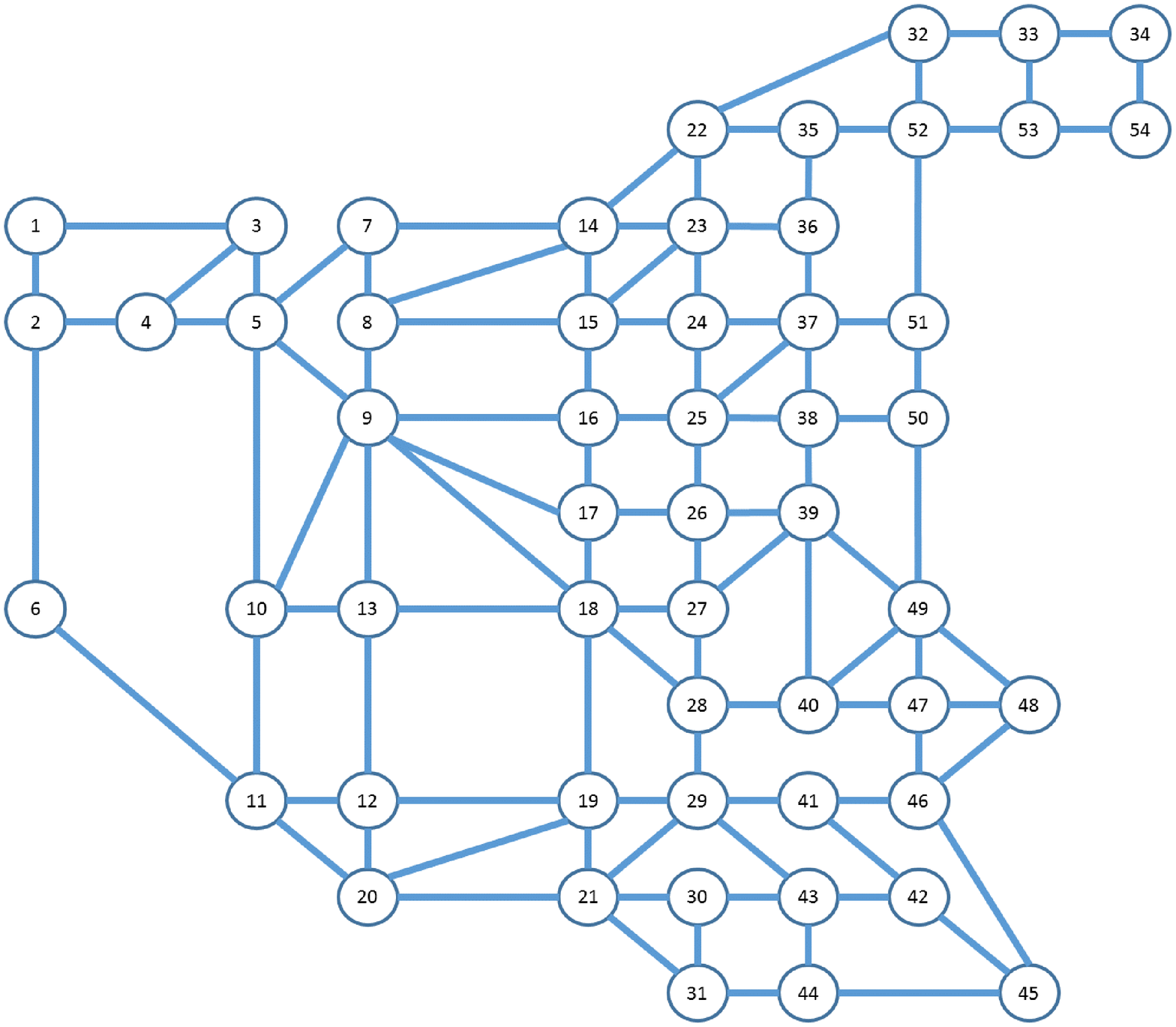}
		\label{fig:Networks:sub:chinese}}
	\caption{Networks used for simulations: (a) NSFNET, 14-Node/22-Link, (b) ARPANET, 20-Node/31-Link, (c) American Backbone\cite{tang2011multicast}, 24-Node/43-Link, and (d) Chinese Backbone\cite{tang2011multicast}, 54-Node/103-Link.}
	\label{fig:Networks}
\end{figure*}

We begin by examining our proposed expansion of redundant quorums by comparing apples-to-apples using the paired cycle routing in prior art.  We used four common networks (Fig. \ref{fig:Networks}) and an implementation of the ECBRA heuristic\cite{dlastine2011ECBRA} to perform the cycle routing.  ECBRA is sensitive to node and edge numbering that a total of 1000 variations on the inputs were considered, each being a one-to-one mapping with the respective network.  For simulation of prior art in \cite{dlastine2014quorum,ckleinheksel2015quorum,ckleinheksel2015redundant}, we used the N=4,...,111 optimal cyclic quorums from \cite{luk1997two}.  Redundant cyclic quorums for $R=2$ and $R=3$ were found using the techniques described in Section \ref{sec:Quorums:sub:Redundant}.

\subsection{Fault-free operational analysis}
It is expected that a majority of the time the optical network will be operating without faults.  It is important that the resource utilization during this period be analyzed.

The metric we use to measure resource utilization is the number of links used in a solution.  Comparing network-to-network is not particularly insightful, but comparing multiple solutions for a particular network is.  The more links that a set of quorum cycles use the fewer (wavelength) resources that can be assigned to each link.  Additionally each logical link represents a required physical transmitter and receiver, hence capital costs.

Table \ref{tbl:paired:links} shows that applying redundancy within the quorums and using paired cycles will lead to an increase in mean network links used (95\% confidence interval.)  $R=1$, column two, is the standard, no redundant pairs, implementation seen in \cite{dlastine2014quorum,ckleinheksel2015quorum}.  $R=2$ and $R=3$ have twice and three times redundant pairs present in their quorum solutions respectively.  Despite having that added redundancy, the resource usage, columns 3 and 4, only increased 5.63-14.18\% and 15.21-22.29\% respectively across the networks.
\begin{table}[t]
	\renewcommand{\arraystretch}{1.3}
	\caption{Mean links used by the paired cycles using our redundant quorum solution (95\% confidence intervals.)}
	\label{tbl:paired:links}
	\centering
	\begin{tabular}{|l|c|c|c|}
		\hline
		Networks & $R=1$             & $R=2$             & $R=3$            \\ \hline
		NSFNET   & 248.96 - 249.88   & 270.21 - 270.93   & 289.95 - 290.85  \\ \hline
		ARPANET  & 510.03 - 511.24   & 538.89 - 539.87   & 587.72 - 588.89  \\ \hline
		American & 641.98 - 643.35   & 718.87 - 720.19   & 752.74 - 753.83  \\ \hline
		Chinese  & 2673.74 - 2678.28 & 3053.24 - 3057.70 & 3270.65 - 3274.51\\ \hline
	\end{tabular}
\end{table}

This resource usage result shows that applying our redundant quorums set technique to paired cycle solutions available today will not pose too significant of a resource burden.  
Next we consider the fault case for paired cycles using our redundant quorum cycle solution and show the increase in resources is being utilized to improve fault recoveries without any optical cycle reconfiguration.

\subsection{Fault-tolerant operational analysis}
Optical networks are highly depended upon.  The fault-tolerance aspect of this route design is critical.  Maintaining the ability to serve all dynamic point-to-point traffic requests despite fault is important.  

We assume fiber link failure(s).  While a simple model, it does cover most real fault scenarios. This occurs when a link is broken, like planned maintenance or the accidental severing during land excavation.  Each modeled node has transmitters and receivers that can fail too.  Short of a natural disaster, devices will likely fail independently of one another.  When a transmitter/receiver pair fails within a modeled node, the effect on the global network is similar to that link failing.  Modeling as a single edge failure, while not an exact fault mapping, is an appropriate abstraction.

\subsubsection{Single fault case}
To model the fault, we simulate the failure of each edge, $e\in E$, in the network model, $G=(V,E)$.  We then examine the network's ability to serve all potential point-to-point requests by counting pairs of nodes that would be able to communicate and conversely those pairs that are unable to communicate.  The results are then reported as fault coverage, i.e. total pairs able to communicate as a percentage of total point-to-point pairs.  100\% would be perfect coverage, whereas 0\% would be no fault coverage at all.

Our simulation results showed our redundant quorum-based cycle technique had 99.83 - 99.98\%  and 99.95 - 99.99\% fault coverage, $R=2$ and $R=3$ respectively, in the four networks tested.  In Table \ref{tbl:paired:singleFault}, we compare the state-of-art paired cycle approach with our redundant technique also with paired cycles.  With single edge failures, the paired cycles had a mean missing communication pair rate of less than 3 pairs or less than 0.53\% across all networks (95\% CI.) Hence column two, it can be seen that the fault coverage is greater than 99.47\%.  Our redundant quorum cycles technique, columns 3 and 4, had a mean missing pair rate (95\% CI) of less than 0.48 and 0.26 respectively across all networks, which is reflected in fault coverages greater than 99.83\%.
\begin{table}[t]
	\renewcommand{\arraystretch}{1.3}
	\caption{Percent mean fault coverage (95\% CI) of paired cycles using our redundant quorum solution experiencing a single link fault.}
	\label{tbl:paired:singleFault}
	\centering
	\begin{tabular}{|l|c|c|c|}
		\hline
		Networks & $R=1$           & $R=2$           & $R=3$           \\ \hline
		NSFNET   & 99.47 - 99.50\% & 99.83 - 99.84\% & 99.99 - 99.99\% \\ \hline
		ARPANET  & 99.80 - 99.80\% & 99.93 - 99.94\% & 99.97 - 99.98\% \\ \hline
		American & 99.62 - 99.63\% & 99.91 - 99.92\% & 99.95 - 99.96\% \\ \hline
		Chinese  & 99.90 - 99.90\% & 99.98 - 99.98\% & 99.99 - 99.99\% \\ \hline
	\end{tabular}
\end{table}

Depending on the network, the difference between a 99.5\% and 99.99\% fault coverage could be significant.  Being able to achieve that with only the moderate overheads examined in the previous section is just one of the benefits of the redundant quorums set technique.  Being able to dial in on the fault coverage desired using single ($R=1$), double ($R=2$), or triple ($R=3$) redundancy also adds to the flexibility.

\subsubsection{Two fault case}
A significantly more complex model considers two faults simultaneously.  All possible two edge failure combinations in the network are simulated.  We then examine the network's ability to serve all potential point-to-point requests by counting pairs of nodes that would be able to communicate and conversely those pairs that are unable to communicate.

\begin{figure}[t]
	\centering
	\includegraphics[width=3.5in]{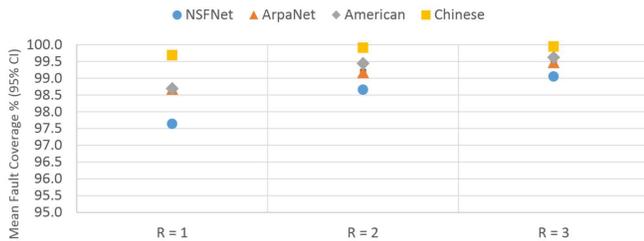}
	\caption{Percent mean fault coverage (95\% CI) of paired cycles using our redundant quorum solution experiencing two simultaneous link faults.  Confidence intervals are shown for each data point, however tight intervals may be hidden by the data point marker.}
	\label{fig:Paired:TwoFault}
\end{figure}

Our simulation results showed our redundant quorum-based cycle technique had 98.65 - 99.91\%  and 99.04 - 99.95\% fault coverage, $R=2$ and $R=3$ respectively, in the four networks tested.  
In Figure \ref{fig:Paired:TwoFault}, we compare the state-of-art paired cycle approach with our redundant technique also with paired cycles.  With two edge failures, the paired cycles had a mean missing communication pair rate of 9 pairs (2.37\%) or less across all networks (95\% CI.) Hence in the $R=1$ column, it can be seen that the fault coverage is 97.63\% or more.  Our redundant quorums set technique, $R=2$ and $R=3$, had a missing pair rate (95\% CI) of 3.14 and 2.09 or less respectively across all networks, which is reflected in fault coverages of 98.65\% or more.

An interesting aspect of the data is that the benefit of additional quorum redundancy in paired cycle solutions is network dependent.  All networks had improved fault coverage with additional quorums set redundancy (Fig. \ref{fig:Paired:TwoFault}), however the NSFNET network had larger increases than others.  Additionally, all networks increasing to $R=3$ had diminishing returns compared to the increases seen when moving from $R=1$ to $R=2$.  

\section{Single Cycle Network Analysis}
\label{sec:singleCycleAnalysis}


In the previous analysis using paired cycle routing (Section \ref{sec:pairedCycleAnalysis}), our proposed generalization of $R$ redundant quorums sets had moderate increases in resource usage and showed improvements to fault coverage.  This section uses a similar experiment setup, however in contrast we are examining using additional quorum redundancy ($R=3$) with just a single cycle compared to the pair of cycles used in prior art.

\subsection{Fault-free operational analysis}
The more links that a set of quorum cycles uses, the fewer (wavelength) resources that can be assigned to each link.  Additionally each logical link represents a required physical transmitter and receiver, hence capital costs.

Table \ref{tbl:single:links} shows significant 38.85 - 42.39\% resource reduction when using $R=3$ redundancy in quorums over the more traditional, prior art methods of simply using paired cycles.  Using $R=2$ gives even better resource reduction, as shown in Table \ref{tbl:single:links} and \cite{ckleinheksel2015redundant}.  This reduction represents the potential for lower capital costs in terms of physical transmitters and receivers needed and the potential for more (wavelength) resource availability within the network.  The paired cycles results with a 95\% confidence interval (CI) for $R=1$ in Table \ref{tbl:paired:links} is repeated in column two of Table \ref{tbl:single:links} for comparison to the single cycle, increased quorum redundancy technique.  Our redundant quorum technique uses far fewer links (shown in columns 3 - 6.)  $R=2$ comes close to halving the necessary resources, whereas $R=3$ is slightly larger at approximately 60\% of the paired $R=1$ solution.
\begin{table*}[t]
	\renewcommand{\arraystretch}{1.3}
	\caption{Mean links used by single cycles compared to paired cycles using our redundant quorum solution (95\% CI.)}
	\label{tbl:single:links}
	\centering
	\begin{tabular}{|l|c|c|r|c|r|}
		\hline
		& $R=1$ (Paired)    & \multicolumn{2}{c|}{$R=2$ (Single)} & \multicolumn{2}{c|}{$R=3$ (Single)} \\ \cline{2-6} 
		Networks & Links             & Links                & Reduction    & Links                & Reduction    \\ \hline
		NSFNET   & 248.96 - 249.88   & 135.10 - 135.46      & -45.76\%     & 144.98 - 145.43      & -41.79\%     \\ \hline
		ARPANET  & 510.03 - 511.24   & 269.44 - 269.93      & -47.19\%     & 293.86 - 294.44      & -42.39\%     \\ \hline
		American & 641.98 - 643.35   & 359.44 - 360.10      & -44.02\%     & 376.37 - 376.92      & -41.39\%     \\ \hline
		Chinese  & 2673.74 - 2678.28 & 1526.62 - 1528.85    & -42.91\%     & 1635.32 - 1637.26    & -38.85\%     \\ \hline
	\end{tabular}
\end{table*}

Previously paired light-trails were used to form all of the point-to-point communication node pairs with minimum sized quorum cycles.  This work and \cite{ckleinheksel2015redundant} consider utilizing intentionally formed redundant node pairs within the quorum routing to reduce the resources used as a potential trade-off to a small cost to network performance.  We analyze the impact of increasing the redundancy within quorums to $R=3$ and its impact of keeping resource utilization low.  To measure this cost, metrics of missing node pairs is used.  

Ideally, like the paired cycle case, there would be 0\% missing, however single cycles don't have the benefit of both $(a_i,a_j)$ and $(a_j,a_i)$ pairs occurring in the same cycle.  Table \ref{tbl:single:missing} shows two important results.  First, the dramatic reduction in resource utilization came at a trade off of a few missing communication pairs.  $R=2$ missed 0.90\% or fewer on average (95\% CI), and $R=3$ missed even fewer at 0.26\% or less on average (95\% CI.)  The paired cycles (column 2, Table \ref{tbl:single:links}) used significantly more resources and did not miss any pairs (column 2, Table \ref{tbl:single:missing}.)  Secondly, compared to $R=2$ single cycles, our redundant $R=3$ cycles performs approximately 2+ times better every time.  As seen in Table \ref{tbl:single:links}, this performance improvement came at a only a slightly higher cost, while still being significantly smaller than the state of art approach.
\begin{table}[t]
	\renewcommand{\arraystretch}{1.3}
	\caption{Mean missing node pairs by single cycles using our redundant quorum solution (95\% CI.)}
	\label{tbl:single:missing}
	\centering
	\begin{tabular}{|l|c|c|c|}
		\hline
		Networks & $R=1$ (Paired) & $R=2$ (Single) & $R=3$ (Single) \\ \hline
		NSFNET   & 0.00\%         & 0.81 - 0.90\%  & 0.03 - 0.04\%  \\ \hline
		ARPANET  & 0.00\%         & 0.33 - 0.37\%  & 0.10 - 0.13\%  \\ \hline
		American & 0.00\%         & 0.50 - 0.54\%  & 0.23 - 0.26\%  \\ \hline
		Chinese  & 0.00\%         & 0.26 - 0.27\%  & 0.08 - 0.09\%  \\ \hline
	\end{tabular}
\end{table}

The quorums set method guarantees that all of the pairs exist (Section \ref{sec:Quorums:sub:Redundant})  It is the limitations of an unidirectional optical light-trails with its required one optical shutter in the off state per cycle that has caused the missing pairs and the potential need for additional compensation steps.  Compensation is possible using an off-the-shelf solution of an additional routing step involving an Optical-to-Electrical-to-Optical (O/E/O) conversion and resending by a hub node.  Even so, on average the $R=2$ and $R=3$ redundant quorums cycle solutions would require infrequent additional steps.

\subsection{Fault-tolerant operational analysis}
Using our generalized $R$ quorum redundancy rather than cycle redundancy can save significant resources, however this cannot come at significant determent to fault tolerance though.

\subsubsection{Single fault case}
Again to model the fault, we simulate the failure of each edge, $e\in E$, in the 1000 node mappings of each network model, $G=(V,E)$.  We then examine the network's ability to serve all potential point-to-point requests by counting pairs of nodes that would be able to communicate and conversely those pairs that are unable to communicate.  The results are then reported as fault coverage, total pairs able to communicate as a percentage of total point-to-point pairs.  100\% would be perfect coverage, whereas 0\% would be no fault coverage at all.

\begin{figure}[t]
	\centering
	\includegraphics[width=3.5in]{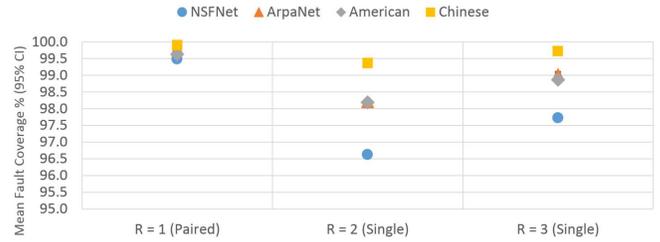}
	\caption{Percent mean fault coverage (95\% CI) of our single cycle, redundant quorum solution experiencing a single link fault.}
	\label{fig:Single:OneFault}
\end{figure}

Our simulation results showed our redundant quorum-based cycle technique had 96.60 - 99.37\%  and 97.70 - 99.72\% fault coverages, $R=2$ and $R=3$ respectively, in the four networks tested.  In Figure \ref{fig:Single:OneFault}, we compare the state-of-art paired cycle approach with our quorum redundant technique with single cycles that uses significantly fewer resources.  With single edge failures, the paired cycles had a mean missing communication pair rate of less than 3 pairs or less than 0.53\% across all networks (95\% CI.) Hence the $R=1$ (Paired) column shows mean fault coverage percentages is greater than 99.47\% for all four networks.  Our redundant quorum cycles technique, $R=2$ and $R=3$ (Single), could not reach that level of coverage, but did achieve an acceptable mean fault coverage rate (95\% CI) of greater than 96.60 and 97.70\% respectively across all networks.

While neither single cycle $R=2$ or $R=3$ could achieve the same level of fault coverage as the paired cycle solution, they did have missing rates better than 3.40 and 2.30\% respectively, while achieving significant resource savings.  In networks where an additional approximately 40\% of resources could better be utilized for communication rather than redundancy, the trade off of missing a relatively small percentage of communications during fault conditions may be considered tolerable.

\subsubsection{Two fault case}
The more complex two fault model considers all possible two edge failure combinations in the simulated networks.  We then examine the network's ability to serve all potential point-to-point requests by counting pairs of nodes that would be able to communicate and conversely those pairs that are unable to communicate.

\begin{figure}[t]
	\centering
	\includegraphics[width=3.5in]{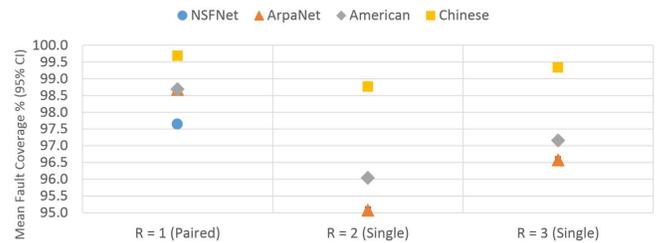}
	\caption{Percent mean fault coverage (95\% CI) of our single cycle, redundant quorum solution experiencing two simultaneous link faults.  For graph clarity and consistency NSFNET's confidence intervals for $R=2$ (Single) at 92.01 - 92.05\% and for $R=3$ (Single) at 93.23 - 93.27\% mean fault coverage were not included in the graph.}
	\label{fig:Single:TwoFault}
\end{figure}

Our simulation results showed our redundant quorum-based cycle technique had 92.01 - 98.77\%  and 93.23 - 99.34\% fault coverage, $R=2$ and $R=3$ respectively, in the four networks tested.  In Figure \ref{fig:Single:TwoFault}, we compare the state-of-art paired cycle approach with our quorum redundant technique with single cycles that uses significantly fewer resources.  With two edge failures, the paired cycles had a mean missing communication pair rate of less than 9 pairs or less than 2.37\% across all networks (95\% CI.) Hence in the $R=1$ (Paired) column, it can be seen that the fault coverage is greater than 97.63\%.  Our redundant quorums set technique, $R=2$ and $R=3$ (Single), could not reach that level for all networks.  Overall the missing pair rate (95\% CI) was less than 7.99 and 6.77\% respectively across all networks, which is reflected in fault coverages greater than 92.01\%.

First, it is worth pointing out again that the paired cycles solution uses more than 38\% more resources on average (95\% CI), while on average the redundant quorums cycle technique using only single cycles performed at most 6\% worse in terms of mean fault coverage on the rarer two simultaneous fault case.  This could be an acceptable trade off in many networks.  Additionally, an interesting aspect of the data is that the results appear to have some network dependence.  The networks with the larger number of nodes had better resiliency to faults than the smaller networks.  This is likely a byproduct of our quorums set solution, where every cycle contains only a small, size $K$, subset of the total nodes.  In all graphs, this translates into cycles that potentially can span the diameter of the graph.  In larger graphs this could lead to more non-quorum nodes being passed through while forming the cycle, which could translate into more than the minimum calculated quorums set pairs (see Sections \ref{sec:Quorums:sub:Definition} and \ref{sec:Quorums:sub:Redundant}.)

\section{Conclusion}
In this paper we proposed and evaluated a generalized formulation of redundant quorums sets for optical cycle routing.  When the generalized $R$ redundancy was applied to state of the art paired cycle techniques it provided optical networks almost fault-tolerant communications for both one and two simultaneous fault cases. When applied to our single cycle routing technique, the $R$ redundant quorums significantly reduced resource usage and maintained high fault tolerance capabilities.

Quorums sets were chosen such that all network communication pairs appeared $R$ times within a routed cycle set.  We intentionally utilized this generalized $R$ redundancy within the quorum cycles for fault-tolerance and reduction in resource usage.  For $R=2$ and $R=3$, the paired cycle techniques on average used 5.63 - 14.18\% and 15.21 - 22.29\% more resources respectively, while on average achieving near fault-tolerance with 98.65 - 99.91\% and 99.04 - 99.95\% fault coverage rates respectively on the two simultaneous faults simulation.  The single cycle technique had similar successes with almost fault-tolerant cycles that used significantly fewer resources (42.91 - 47.19\% and 38.85 - 42.39\% fewer respectively), while at the same time maintaining a high degree of fault-tolerance with 92.01 - 98.77\% and 93.23 - 99.34\% fault coverage respectively on the two simultaneous faults simulation.  

In future work, we are examining ways to improve both the fault and fault-free performance further, while maintaining the significant resource usage reductions that the generalized $R$ redundant quorums cycle solution provides.

\section*{Acknowledgment}
Research funded in part by NSF Graduate Research Fellowship Program, IBM Ph.D. Fellowship Program, Symbi GK-12 Fellowship at Iowa State University, and the Virginia and Phillip Sproul Professorship at Iowa State University.  The research reported in this paper is partially supported by the HPC@ISU equipment at Iowa State University, some of which has been purchased through funding provided by NSF under MRI grant number CNS 1229081 and CRI grant number 1205413.  Any opinions, findings, and conclusions or recommendations expressed in this material are those of the author(s) and do not reflect the views of the funding agencies.


\bibliographystyle{IEEEtran}
\bibliography{../latex_bibs/mybib}

\end{document}